\documentstyle[12pt]{article}

\newcommand\beq{\begin{equation}}
\newcommand\eeq{\end{equation}}
\def\beqa{\begin{eqnarray}}
\def\eeqa{\end{eqnarray}}
\def\bega{\begin{array}}
\def\enda{\end{array}}
\def\be{\[}
\def\ee{\]}
\def\non{{\nonumber}}

\def\l{{\langle}}
\def\r{{\rangle}}

\def\x{{\bf x}}
\def\k{{\bf k}}

\def\eps{\epsilon}

\def\w{\omega}

\def\d#1{{\buildrel \leftrightarrow \over \partial_{#1}}}

\begin{document}

\title{The use of $\exp(iS[x])$\\ in the sum over histories}
\author{Arlen Anderson\thanks{arley@ic.ac.uk}\\
Blackett Laboratory\\
Imperial College\\
Prince Consort Road\\
London SW7 2BZ England}
\date{Aug. 13, 1993}
\maketitle

\vspace{-10cm}
\hfill Imperial-TP-92-93-46

\hfill gr-qc/9308015
\vspace{10cm}

\begin{abstract}
The use of $\sum \exp(iS[x])$ as the generic form for a sum over histories
in configuration space is discussed critically and placed in its proper
context. The standard derivation of the sum over paths by discretizing the
paths is reviewed, and it is shown that the form $\sum \exp(iS[x])$ is
justified only for Schrodinger-type systems which are at most second order
in the momenta. Extending this derivation to the relativistic free
particle, the causal Green's function is expressed as a sum over timelike
paths, and the Feynman Green's function is expressed both as a sum over
paths which only go one way in time and as a sum over paths which move
forward and backward in time. The weighting of the paths is shown not to be
$\exp(iS[x])$ in any of these cases.  The role of the inner product and
the operator ordering of the wave equation in defining the sum over
histories is discussed.
\end{abstract}
\newpage

\baselineskip=24pt

Hartle has made the suggestion that the sum over histories is more
fundamental than canonical quantization and that it may be defined outside
of the Hilbert space context\cite{Har}. This is a provocative proposal
which deserves close investigation. The question of whether there is a
principle which allows one to directly formulate a quantum theory as a sum
over histories has been raised before (see, e.g., \cite{HaK}). Working
within a Hilbert space, this paper discusses and places in its proper
context the traditional view that all path integrals have the schematic
form
\beq
\label{eS}
\l \x'',t''| \x', t'\r= \sum_{x\in{\rm paths}} e^{iS[x]},
\eeq
where $S[x]$ is the action along the path $x$ which begins
from $\x'$ at time $t'$ and ends at $\x''$ at time $t''$.

The sum over histories takes this form in non-relativistic quantum
mechanics and conventional relativistic quantum field theory. In other
contexts, however, while it may be a useful heuristic to guide construction
of a sum over histories quantization, it is not a general principle. In
this paper, path integrals for the causal and Feynman Green's functions for
the free relativistic particle are constructed and shown not to have this
form. The significance of this is that in seeking a general formulation of
the sum over histories, one must look beyond the form (\ref{eS}). In
addition, it is emphasized that the standard definition of the sum over
histories, in terms of the limit of a discretization of paths, relies on
intimate details of the canonical Hilbert space formalism, in particular
upon the operator ordering of the Hamiltonian and the form of the inner
product. This raises several issues which must be addressed in attempting
to define the sum over histories outside of a Hilbert space context.

A recent paper\cite{HaO} shows how the composition laws for several of the
relativistic free particle Green's functions can be derived from the sum
over histories and argues that this is a necessary condition for the
existence of an equivalent canonical formulation. The causal Green's
function is not treated because the authors could not
find a sum over histories representation for it
in configuration space.
They mention that there is a phase space sum over histories but
dismiss it because it does not lead to a sum of the form (\ref{eS}).
As will be shown explicitly, this is a
failing of that form, not an indication that the causal Green's
function cannot be represented as a sum over histories in configuration
space.

In motivating their derivation of the composition laws from the sum over
histories, the authors of
Ref.~\cite{HaO} show that if the sum has the form (\ref{eS}), and if all the
paths travel only forward in time, then the propagator
will satisfy the composition law
\beq
\l x'',t''| x',t' \r = \int d{x_t} \l x'',t''| x_t,t \r \l x_t,t | x',t' \r.
\eeq
{}From this, one can infer that the resolution of
the identity is
\beq
\label{nrelroi}
1=\int d{x_t} | x_t,t \r \l x_t,t |.
\eeq
In contrast, the resolution of the identity for Lorentz-invariant
position eigenstates of the free
relativistic particle on a constant $t$ hypersurface is
\beq
\label{relroi}
1=i\int d{x_t} | x_t,t \r \d0 \l x_t,t |.
\eeq
If this result is to be derived from a sum over histories
representation
of the relativistic free particle Green's functions, then
either the sum is not of the form (\ref{eS}) or the paths in
the sum do
not go forward in time, or both. In
Ref.~\cite{HaO}, the paths are taken to go both forward and backwards in
time and the importance of this is emphasized. Ultimately, the sum used there
is not of
the form (\ref{eS}), but this goes unrecognized, and its significance is
missed.  As will be seen below, the Green's
functions can also be represented as a sum over paths which only move one
direction in time, but the form is not (\ref{eS}).

A second recent paper\cite{ReS} discusses the Newton-Wigner
propagator\cite{HaK} in configuration space.  After finding a path integral
representation, discussed again below, in terms of an infinite product of
Bessel functions, the authors twist their result trying to force it into the
form (\ref{eS}).  They also attempt to make contact with the naive
formal representation of the form (\ref{eS}).  They experience difficulty which
they attribute to differences in the short-time ($\Delta t\rightarrow 0$)
and $\hbar\rightarrow 0$ limits of the propagator.
Properly understood, their computation instead shows the (unmodified)
formal representation is wrong.

Each of these papers suffers from an uncritical commitment to the schem\-atic
form $\sum e^{iS}$.  Generically, though not informatively, any sum over
paths has the form
\beq
\label{sopF}
\sum_{x\in {\rm paths}} F[x],
\eeq
where $F[x]$ is the weight given to the path $x$. As will be reviewed
shortly, the weighting $F[x]=\exp(iS[x])$ holds for parabolic wave
equations which are at most second order in derivatives---that is,
Schrodinger-like equations at most quadratic in the momenta. This form
applies to non-relativistic quantum mechanics and conventional relativistic
quantum field theory (which is equivalent to a functional Schrodinger
equation\cite{Hat}). Out of all possible functional integrals, however,
it is a special form, and in particular it is not appropriate for the
wave equation for the free relativistic particle.

Before studying the relativistic particle, it is instructive to review
the origin of the form (\ref{eS}) in the non-relativistic case.
Recall a standard derivation of the path integral for the
non-relativistic particle\cite{Hat}.  One inserts a sequence of resolutions
of the
identity (\ref{nrelroi}) into the full transition amplitude, breaking it
up into a product of short-time propagators:
\beqa
\label{nrprop}
\l x'',t''| x',t' \r &=& \\
&& \hspace{-3cm} \lim_{N\rightarrow \infty} \int \prod_{n=1}^{N-1}
dx_n \l x'',t''| x_{N-1},t_{N-1} \r
\l x_{N-1},t_{N-1}| x_{N-2},t_{N-2} \r \cdots \l x_1,t_1| x',t' \r \non
\eeqa
The short-time propagator for a Schrodinger equation with Weyl-ordered
Hamiltonian\cite{Weyl} $H(\hat p, \hat q)$
is given by
\beqa
\label{stprop}
\l x_{n+1},t_{n+1}| x_{n},t_{n} \r &=&
\l x_{n+1}| e^{-iH(\hat p,\hat q)\eps}| x_{n} \r \\
&\approx& \l x_{n+1}| (1-i H(\hat p,\hat q) \eps) | x_n \r \non \\
&\approx& \int {dk_n\over 2\pi}
\exp[ik_n(x_{n+1}-x_n) -iH(k_n,{x_{n+1}+x_{n}\over 2}) \eps], \non
\eeqa
where $t_{n+1}-t_{n}\equiv \eps=(t''-t')/N$.
Note that the short-time propagator is only accurate to order $\eps$, but
this is all that is needed\cite{iden}.

For
Hamiltonians which are quadratic in the momenta, the $dk_n$ integral can
be done to give the discretized form of the action.  For example,
for $H(\hat p, \hat q) =  \hat p^2 +V(\hat q)$, this is
\beq
\label{quadH}
\l x_{n+1},t_{n+1}| x_{n},t_{n} \r =
(4\pi i\eps)^{-1/2}\exp[i{(x_{n+1}-x_{n})^2\over  4\eps} -i
V({x_{n+1}+x_{n} \over 2})\eps] .
\eeq
Substituting (\ref{stprop}) into (\ref{nrprop}), one has the discretized
phase space path integral
\beqa
\label{psprop}
\l x'',t''| x',t' \r &=& \\
&&\hspace{-3cm} = \lim_{N\rightarrow \infty} \int \prod_{n=1}^{N-1}
dx_n \prod_{n=0}^{N-1} {dk_n\over 2\pi}
\exp[i\sum_{n=0}^{N-1} k_n(x_{n+1}-x_n) -H(k_n,{x_{n+1}+x_{n}\over 2})
\eps]. \non
\eeqa
For the quadratic Hamiltonian above, using (\ref{quadH}) in (\ref{nrprop}),
and taking the limit so that the sum in the
exponent becomes an integral, one has
\beq
\l x'',t''| x',t' \r = \int {\cal D}x \exp(i\int_{t'}^{t''} {1\over 4}
\dot x^2 -V(x) dt),
\eeq
which has the familiar form (\ref{eS}).

If the Hamiltonian were not quadratic in the momentum, one would still
have a sum over paths, but it would not be of the form (\ref{eS}).
This happens for instance with the Newton-Wigner propagator for a free
relativistic particle, where
$H(\hat p, \hat q)= (\hat p^2 + m^2)^{1/2}$.  The phase space form
(\ref{psprop}) is still valid\cite{HaK}.  If one insists on doing the momentum
integrals, which should be done with care since the integral defines a
distribution\cite{Ful}, one finds that the weighting given to each path
is essentially a product of Bessel functions along the path\cite{ReS}.
Since
Bessel functions do not share the nice property of exponentials that, when
forming the product, the
arguments add, one does not find a simple expression for the infinite
product.

In Ref.~\cite{ReS}, the authors seem to
miss the point of their calculation.  There is no reason,
besides wishful thinking, to believe that the  Newton-Wigner propagator
should have the form $\int {\cal D}x e^{iS[x]}$,
where $S=-m\int (1-\dot x^2)^{1/2} dt$
is the classical action appropriate to the relativistic particle. If it
did, and if the collection of paths summed over were those defined by the
discretization argument above, then one would have to find that for
short-times the Bessel function form of the propagator equals, to order
$\eps$, the exponential of the discrete action
\beq
\exp(-im\eps(1-{(x_{n+1}-x_n)^2\over \eps^2})^{1/2})
\eeq
(up to an overall function of $\eps$).  This isn't possible because while
the $\eps
\rightarrow 0$ limit of the Bessel function form of the propagator is
$\delta(x_{n+1}-x_n)$ as it should be, this other is not!  This is not a
mystery of the relation between the short-time and WKB
($\hbar\rightarrow 0$) approximations as the authors of Ref.~\cite{ReS}
suggest; this is proof that the (unmodified) sum over
$\exp(iS[x])$ is wrong.

The sum over paths expression of the causal Green's function
$iG(x'',x')=\l x''| x'\r$
may now be
constructed [where $x$ now stands for the four-vector $(x^0,{\bf x})$].
One follows the procedure for the non-relativistic case
except one uses the resolution of the identity (\ref{relroi}) appropriate
to the relativistic particle.  One obtains
\beq
\label{cprop}
\l x''| x'\r = \lim_{N\rightarrow \infty} \int \prod_{n=1}^{N-1}
i d^3\x_n \l x''| x_{N-1} \r \d{x^0_{N-1}}
\l x_{N-1}| x_{N-2} \r\d{x^0_{N-2}} \cdots\d{x^0_1} \l x_1| x' \r
\eeq
The short-time (as well as the finite-time) causal propagator is given
by\cite{HaO}
\beq
\label{stcprop}
\l x_{n+1} | x_n \r = {-i\over (2\pi)^3} \int {d^3 {\k_n} \over \w_{\k_n}}
\sin [\w_{\k_n}(x^0_{n+1}-x^0_n)] e^{i\k_n\cdot (\x_{n+1}-\x_n)},
\eeq
where $\w_\k=(\k^2+m^2)^{1/2}$.  The integrand of (\ref{cprop}) is the
weight factor associated to each path.  Because of the Wronskian
derivatives and the sine in the short-time propagator, it is difficult
to express this weight in a compact form.

One may
object that this is a phase space path integral representation, not a
configuration space one\cite{HaO}.  If one wishes, the momentum integrals in
(\ref{stcprop}) may be
evaluated, yielding essentially a Bessel function\cite{BjD}.  This gives
a configuration space integral analogous to that for the Newton-Wigner
propagator in terms of a sum of products of Bessel functions.  As there is
no meaningful reason for preferring the configuration space over
the phase space form once one abandons the mythical $\sum e^{iS[x]}$,
it is not clear why one would insist on doing this.

What is the class of paths which are being summed?  The causal propagator
vanishes for spacelike separated points.  This places a restriction on
the range of each of the $\x_n$ integrations:  if $x^0_{n}-x^0_{n-1}=\eps$,
then the range of the $\x_n$ integration is the ball of radius $\eps$
centered at $\x_{n-1}$.  This means that each step from an $x_{n-1}$
to an $x_n$ must be timelike.  The paths contributing to (\ref{cprop}) are all
the time-like paths between $x'$ and $x''$.  The sum over histories
representation of the causal propagator is thus given by a sum over all
timelike paths between $x'$ and $x''$ weighted by an infinite product of
Wronskian derivatives and
short-time propagators of the form (\ref{stcprop}).

Turn now to one of the other Green's functions of the relativistic
particle,
and consider the Feynman propagator.  This was studied in Ref.~\cite{HaO},
and it was
found that in a proper-time representation the collection of paths summed
over moved both forwards and back in time.  Does this mean that paths
travelling backwards in time are always necessary in constructing
the Feynman propagator?  No, it does not.  Following the construction
just given, one finds the representation for $iG_F(x''|x')=\l x''|x' \r_F$
($x^{0\,\prime
\prime}> x^{0\,\prime}$)
\beqa
\label{fprop}
\l x''| x'\r_F &=& \\
&&\hspace{-2cm} = \lim_{N\rightarrow \infty} \int \prod_{n=1}^{N-1}
i d^3\x_n \l x''| x_{N-1} \r_F \d{x^0_{N-1}}
\l x_{N-1}| x_{N-2} \r_F\d{x^0_{N-2}} \cdots\d{x^0_1} \l x_1| x'
\r_F \non
\eeqa

The composition law for the Feynman propagator involves a
normal deriv\-ative which changes sign if the time-ordering of the
endpoints is reversed.  With a chosen ordering of the endpoints, one
can drop explicit mention of the normal direction. If $x^{0\,\prime \prime}
< x^{0\,\prime}$, one must change the sign of $i$ in the inner product;
this may be achieved by simply taking the complex conjugate of
(\ref{fprop}).  Without the restriction on the relationship of
$x^{0\,\prime \prime}$ and $x^{0\,\prime}$, (\ref{fprop}) is the positive
frequency Wightman function $G^{+}(x''|x')$, and its complex conjugate is
the negative frequency Wightman function $G^{-}(x''|x')$.

The short-time (and finite-time) Feynman propagator is
\beq
\label{fstprop}
\l x_{n+1} | x_n \r_F= i\int {d^4 k\over (2\pi)^4} {e^{-ik\cdot (x_{n+1}
-x_n)}\over k^2 -m^2}.
\eeq
Here, the convention is $k\cdot x=k^0 x^0 -\k\cdot \x$
[$k^2=(k^0)^2-\k^2$].  In the usual way, one defines the contour of the
$k^0$ integration to pass below the pole on the negative
$k^0$ axis and above the pole on the positive axis, as one moves from
negative to positive values of $k^0$.  Alternatively,
one can add $i\alpha$
($\alpha\rightarrow 0$) to the denominator to move the poles in $k^0$ off the
axis.

The Feynman propagator is non-vanishing even for spacelike separations,
so all paths between $x'$ and $x''$ which move forward in $x^0$
(``future-directed'')
contribute to the sum over paths, including those
with spacelike segments.
Paths are not allowed to travel backwards in time.  (If $x^{0\,\prime \prime}
< x^{0\,\prime}$, the sum is over paths travelling only backwards in time.)
One might be concerned about the fact that the specification of
future-directed paths is not Lorentz invariant.  In a second frame
moving relative to the first one, there are future-directed paths
that had appeared to move backwards in time in the first frame.  The point
however is that for every such path one acquires in the new frame, one
loses a previously future-directed path, and there is
always perfect balance.  The detailed collection of paths changes, but
the result is Lorentz invariant.

A test of whether backward-in-time paths necessarily contribute
to the Feynman Green's function is to investigate the composition law for
the case in which the intermediate surface is to the future of both
endpoints.  That is, consider the integral
\beq
\label{comp1}
i \int d^3 \x \l x_2|x \r_F \d{x^0} \l x | x_1 \r_F,
\eeq
where $x^0>x^0_2>x^{0}_1$. One knows that the Feynman Green's function
propagates waves
both forward and back in time, so there is the possibility that one
may take the intermediate surface to the future of both endpoints.  Indeed,
if paths
travelling back in time make a non-trivial contribution, they must show
up here because it is only by backwards travelling paths that the
intermediate surface can influence the final endpoint.

Using the expression
(\ref{fstprop}) for the Feynman propagator, one has
\beq
\label{inp}
- \int d^3 \x {d^4 k_2 d^4 k_1\over (2\pi)^8} {k^0_2+k^0_1 \over
(k_2^2-m^2)(k_1^2 -m^2)} \exp(-ik_2\cdot(x_2-x) -ik_1\cdot
(x-x_1)).
\eeq
Doing the $d^3 \x$ integral gives $(2\pi)^3 \delta^3(\k_2-\k_1)$. The
$d^3\k_1$ integral may be done.  Finally doing the $dk^0_1$ integral by
contour integration, threading between the poles in the usual way for the
Feynman Green's function, and closing the contour in the lower half-plane since
$x^0>x^0_1$, one gets
\beq
\label{comp2}
i\int {d^4 k_2\over (2\pi)^4} {k^0_2+\w_{\k_2}  \over 2\w_{\k_2} (k_2^2
-m^2) } e^{-ik^0_2 (x^0_2-x_0)-i\w_{\k_2}(x^0-x^0_1)+i
\k_2\cdot(\x_2-\x_1)},
\eeq
where $\w_{\k}=(\k^2+m^2)^{1/2}$.
Now, however, one sees that the numerator cancels the pole at
$k^0_2=-\w_{\k_2}$, so when one does the $dk^0_2$ integral and closes the
contour in the upper half-plane, the integral vanishes!  If, instead,
$x^0_2>x^0$, then one would close the integral in the lower half-plane,
and one would reach the correct result.  Thus, one cannot take the
intermediate surface to the future of both endpoints.  The Wronskian
derivative is responsible precisely for removing the pole that allows
backward propagation.  (This calculation can also be read simply as proof
of the orthogonality of the positive and negative frequency Wightman
functions
\be
0=i\int d^3\x G^-(x_2|x) \d{x^0} G^+(x|x_1).)
\ee

What do the authors of Ref.~\cite{HaO} accomplish with their
paths which move both forward and back in time?  By going to the
proper-time representation, they express the Feynman propagator as
\beq
\l x'' | x' \r_F= \int_0^\infty dT \int {d^4 k\over (2\pi)^4}
e^{[-ik\cdot (x''-x')+iT( k^2 -m^2 +i\alpha)]}.
\eeq
The integrand of the $T$ integral is essentially a non-relativistic
free-particle
propagator in the time $T$.  As such, it has a well-known sum over paths
representation following (\ref{nrprop}),
and the infinite product which arises when discretizing it is compactly
represented in terms of the exponential of an action.  This leads to the
sum over paths expression
\beq
\label{pTfprop}
 \l x'' | x' \r_F=\int_0^\infty dT \int \prod_{n=1}^{N-1} { d^4 x_n\over
 i(4\pi T/N)^{2} }
 \exp(-i\sum_{n=0}^{N-1} {(x_{n+1}-x_n)^2\over 4T/N} -i(m^2-i\alpha)T).
\eeq
The sum is over all paths from $x'$ to $x''$ which go forward in $T$,
including paths which go backwards in $x^0$.
No Wronskian derivatives appear, and this is a compact
expression.
The important point is that the paths travelling backwards in time
conspire to implement the effect of the Wronskian derivative. This is what
Ref.~\cite{HaO} explicitly proves.

Eq.~(\ref{pTfprop}) is not of the
form $\sum e^{iS[x]}$.  A path in spacetime is
characterized
by a sequence of points $\{x_n\}$.  The integral over $\prod d^4 x_n$
implements a sum over all paths in spacetime.
Each path in (\ref{pTfprop}) carries an additional parameter,
$T$.  When this parameter is integrated out, the weight associated to the
spacetime path is not of the form $\exp(iS[x])$.  To see this explicitly,
define
\beq
R[x]={N\over 4}\sum_{n=0}^{N-1} (x_{n+1}-x_n)^2.
\eeq
As the sum of the squared proper time separation of the points along the
path, this quantity is the discrete form of some measure of the length of
the path from
$x'=x_0$ to $x''=x_N$.  For the direct path described by the sequence
\beq
\{x_n={n(x''-x')\over N} +x' \},
\eeq
one has $R_{\rm dir}=(x''-x')^2/4$.  The weight for a given path is
given by
\beqa
F[x]&=& \int_0^\infty dT\, T^{2-2N} \exp (-{iR\over T}-im^2 T) \\
&=& \biggl( {\pm R^{1/2}\over im} \biggr)^{3-2N} i\pi H^{(1)}_{2N-3}(\mp 2m
R^{1/2}). \non
\eeqa
Clearly, this is not the exponential of the classical action for the
relativistic particle.

It is interesting to note that the composition law that is obtained from
the computation in Ref.~\cite{HaO} is ($x^0>x^0_1$)
\beq
\l x_2 | x_1 \r_F =2i\int d^3 \x \l x_2|x \r_F \vec \partial_{x^0}
\l x | x_1 \r_F.
\eeq
If this were used in the calculation above, with the intermediate
hypersurface to the future of both endpoints, the numerator in the
expression analogous to (\ref{inp}) would be $2k^0_1$.  After doing the
$\x$ and $\k_1$ integrations, one finds that the pole allowing backwards
propagation is not cancelled.  The integral is non-zero and gives
the correct result.  The backwards moving paths do indeed contribute.
This is nevertheless consistent with the result above.
In Ref.~\cite{HaO} a second composition law is also
obtained which when averaged with the first gives the Wronskian derivative
form of the composition law
and leads to the cancellation of the contributions of the backwards
moving paths.

As a final exercise, it is instructive to verify that one can make the
transformation from the Wronskian to
proper-time representations of the sum over paths for the Feynman
propagator.  This can be done by an iterated manipulation of the composition
expression
\beq
i\int d^3 \x_1 \l x_2|x_1 \r_F \d{x^0_1} \l x_1 | x_0 \r_F,
\eeq
which is the basic building block of the path
integral (\ref{fprop}).  Follow the manipulations above from (\ref{comp1}) to
(\ref{comp2}).  For $x^0_2>x^0_0$, the contour can be closed in the lower
half-plane, and integration gives
\beq
\int {d^3 \k_{1}\over (2\pi)^3} {1  \over 2\w_{\k_1} }
e^{-i\w_{\k_1}(x^0_{2}-x^0_{0})+i \k_1\cdot(\x_{2}-\x_{0})}.
\eeq
With the condition that $x^0_2>x^0_0$, this is equal to
\beq
i\int {d^4 k_1\over (2\pi)^4} {1  \over  k_1^2
-m^2 } e^{-ik^0_1 (x^0_2-x^0_0)+i \k_1\cdot(\x_2-\x_0)}.
\eeq
The purpose of doing and then undoing the $dk^0_1$ contour integration is
to eliminate the dependence on $x^0_1$.  A $d^3\k_0$ and $d^3\x_1$
integration can be reintroduced by inserting and expanding
$\delta^{(3)}(\k_1-\k_0)$
\beq
i\int d^3\x_1 {d^4 k_1\over (2\pi)^4}{d^3\k_0\over (2\pi)^3} {1 \over k_1^2
-m^2 } e^{-ik^0_1 (x^0_2-x^0_0)+i\k_1\cdot(\x_2-\x_1)+i\k_0\cdot(\x_1-\x_0)}.
\eeq

Iteratively applying these manipulations, one finds for the full sum over
paths
\beqa
\l x'',t''| x',t' \r_F &=& \\
&&\hspace{-2cm} = \lim_{N\rightarrow \infty} i \int \prod_{n=1}^{N-1}
d^3 \x_n \prod_{n=0}^{N-1} {d^3 \k_n\over (2\pi)^3} {dk^0_{N-1}\over 2\pi}
{1\over k_{N-1}^2 -m^2} \non \\
&& \hspace{-1cm}
\exp[i\sum_{n=0}^{N-1} \k_n(\x_{n+1}-\x_n) -ik^0_{N-1}(x^0_N-x^0_0)]. \non
\eeqa
Because the $\x_n$ integration implies that all the $\k_n$ are equal,
one can write
\beq
k_{N-1}^2 -m^2=(k^0_{N-1})^2 - {1\over N}\sum_{n=0}^{N-1} \k_n^2 -m^2,
\eeq
and make the replacement
\beq
{1\over k_{N-1}^2 -m^2}=-i\int_0^\infty dT \exp[iT((k^0_{N-1})^2 -
{1\over N}\sum_{n=0}^{N-1} \k_n^2 -m^2)].
\eeq
A $dk^0_n$ and $dx^0_n$ integration can be introduced by inserting
and expanding $\delta(k^0_{n+1}-k^0_n)$.  This produces the desired
proper-time sum over paths
\beqa
\l x'',t''| x',t' \r_F &=& \\
&&\hspace{-2cm} = \lim_{N\rightarrow \infty} \int_0^\infty dT
\int \prod_{n=1}^{N-1}
d^4 x_n \prod_{n=0}^{N-1} {d^4 k_n\over (2\pi)^4} \non \\
&&\hspace{-1cm}
\exp[-i\sum_{n=0}^{N-1} k_n(x_{n+1}-x_n) +{iT\over N}\sum_{n=0}^{N-1}
(k_n^2-m^2)] . \non
\eeqa
Integrating out the momenta gives (\ref{pTfprop}).
This shows that while the presence of Wronskian derivatives in the
expression for the sum over paths is unfamiliar, they can be handled and
indeed are equivalent to other sum over paths representations in which
they do not appear.

The lesson to be learned from the free relativistic particle is that path
integrals in configuration space need not have the form $\sum \exp(iS[x])$.
Using the Hilbert
space structure in the form of the resolution of the identity
(\ref{relroi}), a configuration space sum over paths of a different form
was constructed for the causal
Green's function and the Feynman Green's function.  With a different
resolution of the identity\cite{HaO}, one could go on to construct the
Hadamard Green's
function.  Only paths moving one way in time were
needed to construct the Green's functions.  For the causal Green's
function, only timelike paths contributed, while for the Feynman Green's
function, spacelike paths also contributed.

This procedure of building the sum over histories using insertions of the
resolution of identity can be extended and applied to construct
relativistic
Green's functions in other contexts, e.g. in curved spacetime.  Several
questions arise for further investigation.  Is there a decomposition of
the evolution operator in relativistic systems
like that for nonrelativistic ones discussed in \cite{iden}?  Using this,
is there an analog of the form $\sum e^{iS[x]}$ so that one need not
implicitly know the finite-time propagator to construct the sum over paths?
How are the special requirements for the existence of a postive
frequency decomposition, namely, that a spacetime admit a timelike
Killing vector field, reflected in the construction of the sum over histories?
The causal Green's function is well-defined in any globally hyperbolic
spacetime.  Presumably by the above, this means it always has a sum over
histories representation.  What replaces the other Green's
functions when a positive frequency decomposition is not possible?

An important element in formulating a sum over histories is to define the
class of paths being summed.  In the approach here, this class is obtained
as the limit
of discretized paths.  What happens if one uses a different foliation
for the discretization by inserting resolutions of the identity appropriate
to a different collection of hypersurfaces?  Can one prove that the
quantum theories are equivalent?  This is one way that the problem of
time\cite{Kuc} arises in the sum over histories formulation.

This issue
of precisely defining the class of paths to be summed is one of the central
challenges when
one tries to move outside the Hilbert space setting.  It is easy to write
down formal sums (\ref{sopF}) which do not admit a composition
law.  One can claim then that they cannot be equivalent to any canonical
formulation.  Unfortunately, the theories are not well-defined until
one can precisely specify the class of paths included in the sum.  One of
the functions of the composition law at the present stage of
understanding is to enable characterization of the paths.

This point should be emphasized.  The Hilbert space structure has played
a key role here in the construction of the sum over paths. The inner
product is explicitly used in discretizing the paths.  As well, the
operator ordering of the evolution operator enters in the discretization
process (cf. (\ref{stprop}) and \cite{iden}) and determines the discretized
form of the path weighting.  The inner product and factor ordering of
the wave equation are built into the sum over paths,
through its definition as the limit of a discretization.

When one makes
formal manipulations of the path integral (e.g. \cite{HaHar}),
one must take care not to assume
idealized properties, such as invariance of the measure, which do not
hold when the path integral is more carefully defined.  As well,
``details'' like operator ordering cannot be ignored.
Such assumptions lead to fallacious arguments. (For example, the argument in
\cite{HaHar} would imply that there are no anomalies when a classical
algebra is quantized.) A simple example will make the point.

The Liouville measure in the continuum limit of the
non-relativistic phase space path
integral (\ref{psprop}) naively appears to be invariant under point
canonical transformations $(q,p)\mapsto (f(q), f^{\prime\,-1}p)$. This
would lead one to believe that making this transformation in the phase
space path integral simply involves making the classical transformation in
the phase space action. This would give the wrong result. In general a
point canonical transformation produces an effective potential which must
be added to the classical action\cite{GeJ}. This effective potential
reflects contributions which arise from operator ordering in the operator
Hamiltonian after the transformation is made\cite{AnA}.
In a phase space description,
some of the effective potential arises from the non-invariant
transformation of the discretized Liouville measure\cite{Ger}.
Essentially the non-invariance arises from the necessity of using a
discretization scheme which reflects the operator ordering of the
transformed momentum $f^{\prime\,-1}p$. The lesson that this example
teaches is that a measure which is naively invariant may not be so when it
is carefully defined. (Incidentally, to counter the disingenuous suggestion
that this is solely a problem with the discretized measure, if one were to
choose an operator ordering of
the transformed momentum so that the discretized measure is invariant, the
effective potential still arises from the detailed discretization of the
action. One cannot get away from the need to modify the classical action
because of quantum operator ordering effects.)

As one tries to move beyond the Hilbert space
setting, especially in quantum cosmology, it will be important to strive
to make the sum over histories well-defined, else the work may be subject
to Pauli's epithet of being ``not even wrong.''
It is an important challenge to continue to develop the sum over
histories formulation of quantum mechanics.  Clearly, there are
many interesting questions and much work to be
done.  Having placed the form $\sum e^{iS[x]}$ in its proper context,
we need no longer be constrained by its limitations.

\vspace{1cm}
I would like to thank J.J. Halliwell for discussions on this topic.


\begin{thebibliography}{99}
\bibitem{Weyl} Weyl-ordering is defined by
\be
(\hat q^n \hat p^m)_{\rm Weyl}= 2^{-n} \sum_{k=0}^n \biggl(
{n \atop k} \biggr) \hat q^{n-k} \hat p^m \hat q^k
\ee
\end{thebibliography}
\end{document}